\documentclass[sigconf]{acmart}

\usepackage{amsmath}
\makeatletter
\@ifpackageloaded{newtxmath}{}{\usepackage{amssymb}}
\makeatother
\usepackage{booktabs}
\usepackage{graphicx}
\usepackage{float} 
\usepackage{algorithm}
\usepackage{algorithmic}
\usepackage{balance}
\usepackage{multirow} % Added for multi-level table headers
\usepackage{tabularx}

\AtBeginDocument{%
  }

\copyrightyear{2027}
\acmYear{2027}
\setcopyright{cc}
\setcctype{by}
\acmConference[WSDM '27]{Proceedings of the Twentieth ACM International Conference on Web Search and Data Mining}{February 15--19, 2027}{Hong Kong, Hong Kong}
\acmBooktitle{Proceedings of the Twentieth ACM International Conference on Web Search and Data Mining (WSDM '27), February 15--19, 2027, Hong Kong, Hong Kong}
\acmDOI{10.1145/3862322.3863838}
\acmISBN{979-8-4007-3117-4/2027/02}
\makeatletter
\patchcmd{\@mkbibcitation}{\small}{\small\interlinepenalty=10000\relax}{}{}
\makeatother
\begin{document}

\title{OneQuery: Unified Sequence-to-Structure Generation for Multilingual E-commerce Query Understanding}

\author{Mingyu Zhang}
\authornote{Both authors contributed equally to this research.}
\email{mingyuz@stu.xmu.edu.cn}
\affiliation{%
  \institution{Xiamen University}
  \city{Xiamen}
  \country{China}
}
\author{Jiejie Qin}
\authornotemark[2]
\correspondingauthor
\email{qinjiejie.qjj@alibaba-inc.com}
\affiliation{%
  \institution{Alibaba Group}
  \city{Hangzhou}
  \country{China}
}
\author{Jianhui Ji}
\email{jianhui.jjh@alibaba-inc.com}
\affiliation{%
  \institution{Alibaba Group}
  \city{Hangzhou}
  \country{China}
}
\author{Rong Xiao}
\email{xiaorong.xr@taobao.com}
\affiliation{%
  \institution{Alibaba Group}
  \city{Hangzhou}
  \country{China}
}
\author{Meishan Zhang}
\email{zhangmeishan@hit.edu.cn}
\affiliation{%
  \institution{Harbin Institute of Technology}
  \city{Shenzhen}
  \country{China}
}
\author{Xiaoyi Zeng}
\email{yuanhan@taobao.com}
\affiliation{%
  \institution{Alibaba Group}
  \city{Hangzhou}
  \country{China}
}
\renewcommand{\shortauthors}{Zhang et al.}
%增加了不同大小的sizes都有提升这句话
\begin{abstract}
Multilingual e-commerce search requires robust query understanding to convert cross-lingual user queries with varied linguistic patterns into standardized semantic representations that support retrieval and ranking. However, traditional Query Understanding systems suffer from error accumulation, separately optimized subtasks that ignore their mutual correlations, and high maintenance costs. To address these issues, this paper proposes OneQuery, a unified generative framework for multilingual query understanding based on domain-specific large language models in e-commerce. We reformulate four subtasks, namely Language Recognition, Spelling Correction, Translation, and Query Tagging, into a unified Sequence-to-Structure autoregressive generation process constrained by predefined schemas, and design a three-stage progressive training pipeline with multi-task reinforcement learning alignment. OneQuery demonstrates consistent gains over baselines across all tasks and remains effective across different backbone sizes, and online experiments in the AliExpress e-commerce search scenario further confirm its practical value for industrial deployment.
\end{abstract}



%关键词要在摘要里出现

\maketitle

\section{Introduction}
\label{sec:intro}
Multilingual e-commerce search has become a vital bridge connecting global consumers with a massive, linguistically heterogeneous product supply~\cite{2}. The core difficulty of this scenario lies in its two-sided language complexity: on the supply side, product titles and descriptions are written in a mixture of diverse local languages and English, while on the demand side, users express their shopping intentions in a variety of local languages, mixed-language queries, or misspelled text. Consequently, accurately interpreting such linguistically complex user queries and mapping them into standardized, structured semantic representations becomes crucial to the final search experience. This task, known as Query Understanding (QU), serves as the core upstream hub of the multilingual e-commerce retrieval technology stack. The recent progress of large language models (LLMs)~\cite{25} provides a new opportunity to revisit the design of multilingual e-commerce QU systems~\cite{13, 49}. Recent foundation models, such as DeepSeek-V4~\cite{9}, Qwen3~\cite{10}, and GPT-5~\cite{12}, have demonstrated strong capabilities in natural language understanding, generation, and knowledge-intensive reasoning.

Traditional QU systems typically rely on pipelines composed of multiple independent, task-specific discriminative models~\cite{6, 15, 16, 17, 21, 48}. While such architectures provide modularity, they introduce several practical challenges in production environments. Each subtask requires a separately trained, deployed, and maintained model, which increases computational cost and lengthens the inference pipeline. Since these models are usually optimized independently, they cannot fully capture the dependencies among heterogeneous QU subtasks. As a result, errors from upstream components can propagate to downstream modules and degrade the quality of the final structured query representation. Moreover, frequent model iterations often require re-aligning multiple components maintained by different owners, further increasing maintenance overhead. Motivated by these limitations, recent studies have explored unified generative modeling as a promising direction for jointly handling heterogeneous QU subtasks within a single model~\cite{22, 23, 40, 41}.

To address these challenges, we propose OneQuery, a unified generative framework for query understanding in AliExpress multilingual e-commerce search. OneQuery reformulates four isolated subtasks, namely Language Recognition, Spelling Correction, Translation (into English), and Query Tagging (QT), as a single schema-guided structured generation task. Instead of maintaining separate task-specific models, OneQuery jointly models all subtasks within the shared parameter space of a domain-specific large language model. By generating all fields within the same structured output, OneQuery enables the model to capture dependencies among subtasks within a unified optimization framework~\cite{28}.

The main technical and industrial contributions of this paper are summarized as follows.
\begin{enumerate}
    \item \textbf{Unified Schema-Guided Generation Paradigm.} We propose OneQuery, a unified generative framework for multilingual e-commerce QU. It reformulates Language Recognition, Spelling Correction, Translation, and QT as a schema-guided JSON generation task. By modeling these subtasks within a shared parameter space, OneQuery captures dependencies among heterogeneous QU subtasks, helping alleviate error propagation while reducing the system redundancy and maintenance overhead of serving multiple independent models.

    \item \textbf{Progressive Three-Stage Training and Data Synthesis.} We design a progressive post-training pipeline consisting of Joint Single-Task SFT (Stage 1), Data Synthesis via Iterative Field Completion and Joint OneQuery SFT (Stage 2), and Task-aware RL Alignment (Stage 3). Stage 2 reduces the need for fully annotated multi-task data by synthesizing complete OneQuery training instances from partially annotated task-specific data. Stage 3 further aligns the model with a task-aware multi-objective reward function, where subtask-specific reward designs enable fine-grained optimization across heterogeneous outputs and help reduce conflicts among subtask objectives.

    \item \textbf{Offline Effectiveness and Online Impact.} We evaluate OneQuery through comparative experiments and ablation studies on a test set of 19,087 multilingual search queries from AliExpress. The three-stage pipeline yields consistent gains across 4B, 8B, and 14B backbones, and OneQuery outperforms strong proprietary LLMs prompted in a few-shot manner. Online experiments in the AliExpress search scenario further show positive business impact, including a 0.5\% improvement in L2P and a 0.8-point gain in relevance evaluation, demonstrating the practical value of OneQuery.
\end{enumerate}

\section{Related Work}
Query Understanding (QU) has long been a core component of information retrieval systems. Early multilingual e-commerce QU systems often relied on statistical models, handcrafted rules, and dictionary-based resources for query segmentation, normalization, translation, and tagging~\cite{2}. With the rise of pre-trained language models, BERT~\cite{15}, RoBERTa~\cite{16}, and multilingual variants such as XLM-RoBERTa~\cite{37} have been widely adopted in QU systems. More recently, LLMs have been applied to individual QU subtasks such as query rewriting, intent detection, named entity recognition etc, often through prompting or instruction tuning~\cite{38, 39, 49}. However, existing approaches often formulate QU subtasks in isolation, leaving their dependencies underexplored.

Domain-specific adaptation of LLMs for e-commerce has recently attracted increasing attention. Prior studies adapt LLMs to e-commerce scenarios through instruction tuning, continued pretraining, or reasoning-style modeling for tasks such as product understanding, task generalization, and relevance prediction~\cite{4,5,24,46}. These studies highlight the value of e-commerce domain adaptation. Following this direction, OneQuery builds upon a continually multilingual pretrained e-commerce base model and adapts it to multilingual query understanding with schema-guided structured generation.

Recent studies have explored generative and unified modeling paradigms for industrial search and query understanding. Kuaishou's OneSearch~\cite{23}, inspired by the text-to-text paradigm of T5~\cite{22}, studies end-to-end generative modeling for retrieval and ranking. FoodGPT~\cite{28} constructs instruction data over multiple NLP subtasks with SFT and PPO-based post-training. LinkedIn SQU~\cite{27} adopts schema-constrained generation with a small language model to reduce online latency, and a companion line of work further studies how to scale up LLM serving systems for semantic search~\cite{19}, while Xiaohongshu's QP-OneModel~\cite{26} formulates NER, tokenization, term weighting, and classification as single-stage JSON generation.

Compared with existing unified QU systems~\cite{26,27}, OneQuery differs in three key aspects. First, unlike prior systems that mainly focus on monolingual query understanding, OneQuery is designed for AliExpress multilingual e-commerce search and explicitly incorporates language recognition and query translation as core QU subtasks, which are rarely integrated into existing systems. This enables the model to jointly handle diverse multilingual expressions, spelling variants, and cross-lingual product or brand mentions within a single structured output. Second, building a unified QU model requires training instances with complete annotations for all output fields, which are expensive to obtain in practice. To address this data construction challenge, OneQuery proposes Iterative Field Completion, which converts partially annotated task-specific data into complete structured training instances. Third, while some existing unified QU methods incorporate RL, subtask-specific alignment for heterogeneous subtasks remains less explored. OneQuery further introduces task-aware multi-objective RL alignment, with subtask-specific rewards designed for subtasks.

Recent advances in LLM provide useful tools for data construction, alignment, and knowledge grounding. For data construction, LLM-based data self-synthesis and self-distillation have become common practices for reducing annotation costs and expanding training coverage~\cite{42}. OneQuery follows this line of work and applies it to the practical partial-annotation setting, where task-specific annotations are completed into unified structured outputs. For alignment, InstructGPT popularized the SFT-RM-PPO pipeline~\cite{45}, while PPO\textquoteright s memory overhead~\cite{31} has motivated more efficient alternatives such as GRPO, which removes the value network~\cite{30}. Recent work further shows that rule-based verifiable rewards can be effective without training an additional reward model~\cite{36}, motivating our task-aware rule-based reward design.

\section{Preliminaries}

\subsection{Problem Formulation}
Traditional QU architectures typically involve pipelines composed of independent discriminative models responsible for Language Recognition, Spelling Correction, Query Translation, and Query Tagging, which suffer from the error accumulation and high maintenance cost discussed in Section~\ref{sec:intro}.

To address these issues, we formulate query understanding as a sequence-to-structure generation problem. Specifically, given a user's original query $x$, the goal is to use a single LLM to directly generate a structured JSON object $Y$ containing the outputs of all subtasks. Formally, we define the joint output as $Y = \{y_{\text{lang}}, y_{\text{spell}}, y_{\text{trans}}, y_{\text{qt}}\}$ according to the following schema.
\begin{equation}
Y = \{\, \mathrm{lang}\colon y_{\text{lang}},\ \mathrm{spell}\colon y_{\text{spell}},\ \mathrm{trans}\colon y_{\text{trans}},\ \mathrm{qt}\colon y_{\text{qt}} \,\}
\end{equation}
where $y_{\text{lang}}$ represents language tags, $y_{\text{spell}}$ is the corrected query, $y_{\text{trans}}$ is the translated English query, and $y_{\text{qt}}$ represents structured semantic tags. By generating all fields within a unified structure, this formulation enables the model to share semantic representations across subtasks and model their interactions.

\begin{figure}[t]
  \centering
  \includegraphics[width=\linewidth]{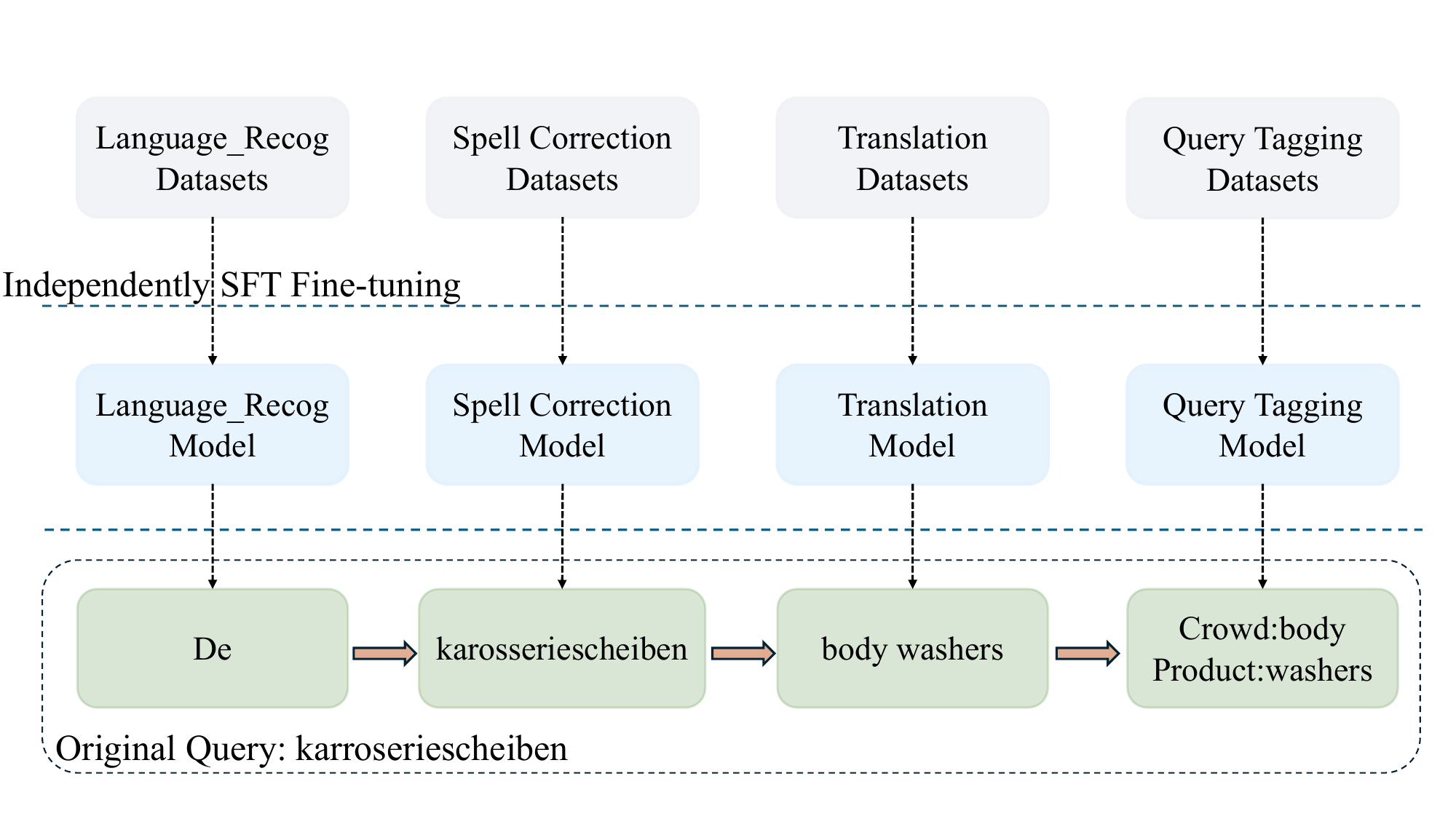}
  \caption{The traditional query understanding pipeline.}
  \label{fig:pipeline}
\end{figure}

%整段新加的
\subsection{Traditional QU Pipeline}
\label{sec:pipeline}
As the baseline of this work, we describe the Traditional QU pipeline currently used in production. The four subtasks are handled by four independent discriminative models, each trained separately on its own annotated dataset with its own objective. Language Recognition is formulated as a sentence-level classification task, Spelling Correction as a sequence rewriting task, Query Translation as a neural machine translation task, and QT as a token-level sequence labeling task over the English query. Since no parameters are shared across these models, each of them is developed, deployed, and iterated independently.

At inference time, the four models are invoked in a fixed order, as illustrated in Fig.~\ref{fig:pipeline}. The original query is first sent to the Language Recognition model. If the query is recognized as a non-English language, it is corrected by the Spelling Correction model and then translated into English by the Query Translation model. The resulting English query is finally passed to the QT model, which produces the structured semantic tags. Each downstream module therefore observes only the output of its upstream module rather than the original query, so an error made at any stage cannot be recovered by the subsequent modules.

\section{The OneQuery Framework}
\begin{figure*}[t]
  \centering
  \includegraphics[width=\textwidth]{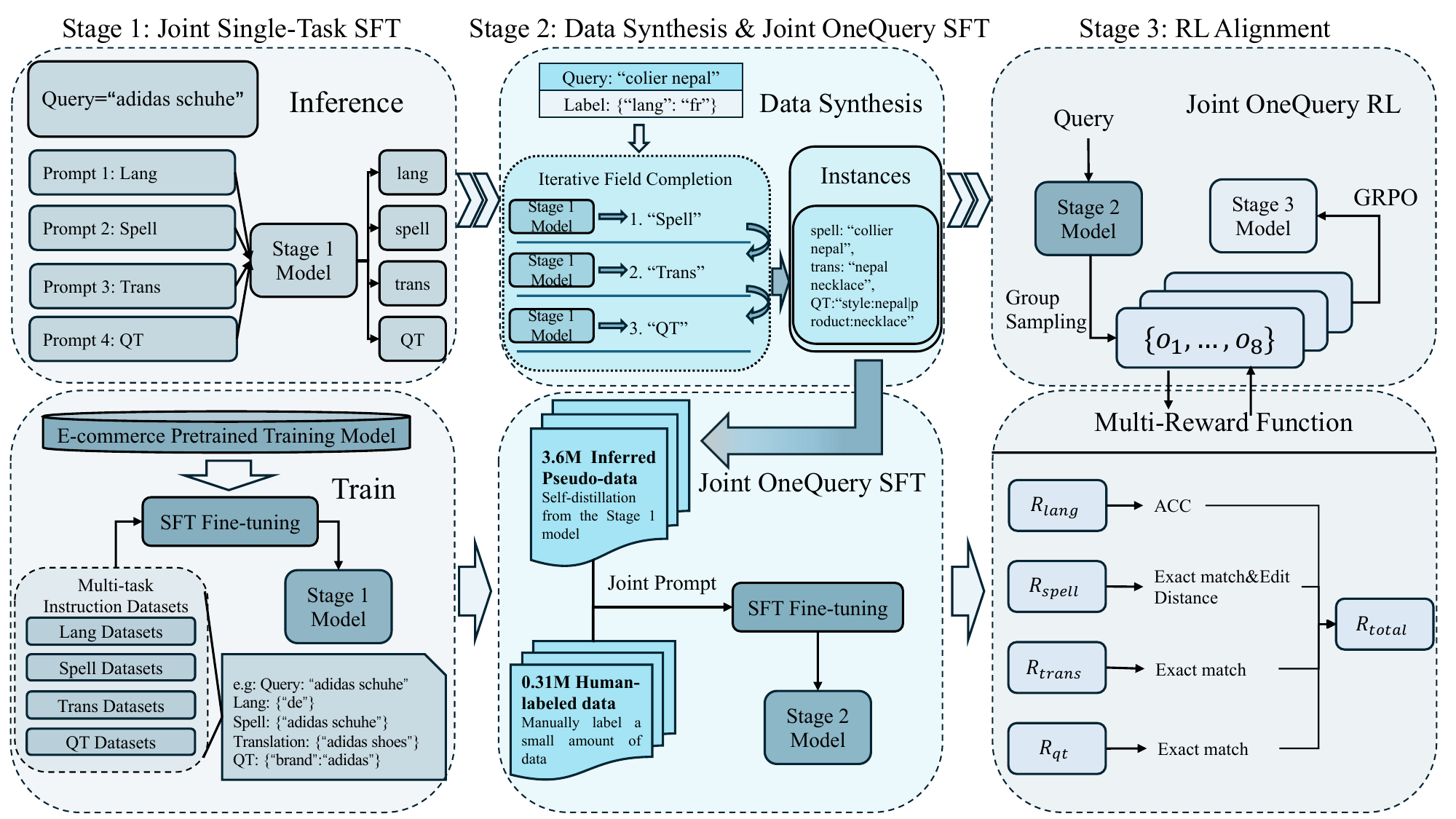}
  \caption{Three-Stage Training Framework of OneQuery.}
  \label{fig:three_stages}
\end{figure*}

To transform a general-purpose LLM into a specialized e-commerce query understanding engine, we design a progressive three-stage post-training pipeline.
\begin{enumerate}
    \item \textbf{Stage 1 (Joint Single-Task SFT).} We mix training samples from the four subtasks, while each sample supervises only one target field. This stage establishes the foundational capability for each individual subtask.
    \item \textbf{Stage 2 (Data Synthesis \& Joint OneQuery SFT).} This stage uses Iterative Field Completion to synthesize $3.6 \times 10^6$ fully annotated samples, enabling the model to learn autoregressive generation of the complete OneQuery JSON structure.
    \item \textbf{Stage 3 (Multi-Task Reinforcement Learning Alignment).} We introduce GRPO~\cite{30} to achieve fine-grained alignment through task-aware multi-reward rules.
\end{enumerate}
Fig.~\ref{fig:three_stages} illustrates the overall framework of the three-stage training pipeline.

\subsection{Stage 1: Joint Single-Task SFT}
\begin{figure}[t]
  \centering
  \includegraphics[width=\linewidth]{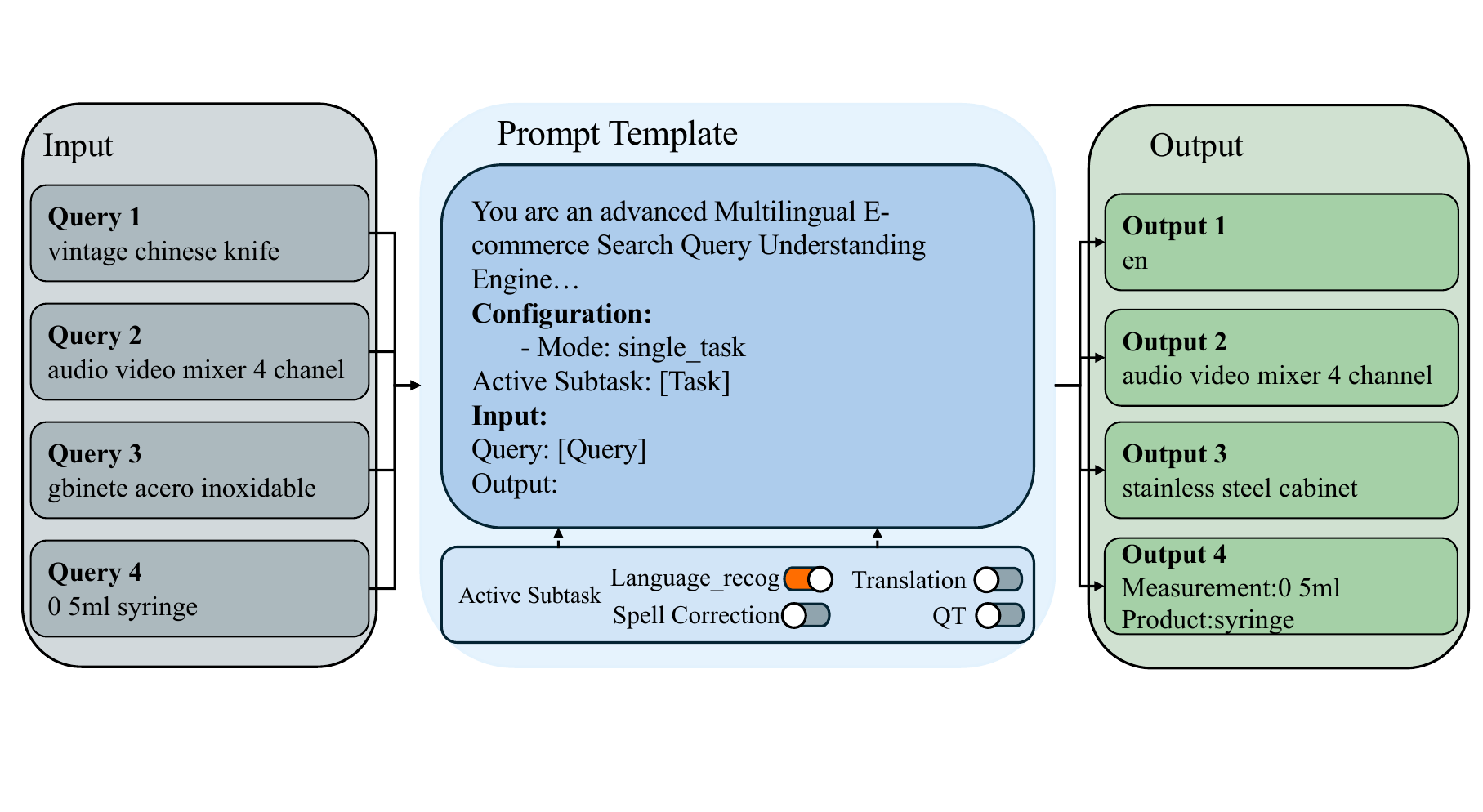}
  \caption{Instruction Prompt Template and Example Outputs for Stage 1 Joint Single-Task SFT.}
  \label{fig:stage1_prompt}
\end{figure}

In Stage 1, we adopt a joint training strategy with independent instruction templates for multi-tasking. We design distinct prompts for each of the four subtasks. As illustrated in Fig.~\ref{fig:stage1_prompt}, these prompts share a global role definition and output format specification, but dynamically specify the task to be performed via an Active Subtask field. The system prompt defines the general task while the instruction defines the specific operation for the model.

For example, when Active Subtask is \textit{Language\_recog}, the model outputs only the language tag; when it is \textit{QT}, the model outputs only the corresponding structured semantic tags. This joint single-task training strategy helps the model acquire strong single-task capabilities before OneQuery training, providing a reliable foundation for subsequent Iterative Field Completion and Joint OneQuery SFT.

\subsection{Stage 2: Data Synthesis \& Joint OneQuery SFT}
Stage 1 addresses capability injection, but the current model still lacks the ability to produce all results in a single generation. Stage 2 aims to train the model for the generation of unified OneQuery results. A key motivation for unification lies in the impact of upstream task outputs on downstream tasks. In traditional pipeline systems, translation accuracy largely determines whether QT produces reliable structured representations, since an erroneous translation is irreversibly passed to the tagging module. 

However, high-quality, comprehensively annotated OneQuery data covering all four fields is extremely scarce and costly to produce. Consequently, we propose a ``Pseudo-data synthesis + Human-labeled'' data construction scheme.

% %%误差累积不应该在这出现，QT对但是trans错的case，放在主实验下面的一个小章节

% \begin{figure}[H]
%   \centering
%   \includegraphics[width=\columnwidth]{trans-qt-propag}
%   \caption{
%     Critical Impact of Translation Accuracy on Query Tagging
%   }
%   \label{fig:trans-qt-propag}
% \end{figure}

\subsubsection{OneQuery Data Synthesis Pipeline}
The model trained in Stage 1 learns task-specific capabilities for language recognition, spelling correction, query translation, and QT through task-specific instruction tuning. However, since each training instance supervises only one target field, the model has not yet learned to combine these capabilities to generate the complete OneQuery JSON structure.

A key challenge is that the Stage 1 datasets are task-specific and partially annotated. For a given query $x$, we may only have its language label $y_{\text{lang}}$, while annotations for spelling correction, query translation, or QT are missing. Therefore, we cannot directly build complete OneQuery training samples from the existing task-specific annotations.

To address this data construction challenge, we design an Iterative Field Completion strategy. It leverages the trained Stage 1 model as a generative annotator to progressively fill in missing fields over large-scale query data. Specifically, for any query $x$, the target output structure is defined as $Y=\{y_{\text{lang}}, y_{\text{spell}}, y_{\text{trans}}, y_{\text{qt}}\}$. We sequentially construct a task-specific instruction prompt $P_k$ for each field $y_k$ to be completed, where $k \in \{\text{lang, spell, trans, qt}\}$. We set the Active Subtask within $P_k$ to its corresponding task name and perform inference upon $x$.
\begin{equation}
\hat{y}_k = \mathrm{Decode}\left(\mathcal{M}_{\text{stage1}}, P_k, x\right)
\end{equation}

Through this iterative generation process, the model supplements all missing fields for query $x$. During final aggregation, we prioritize assembling existing high-quality ground-truth labels $y_k^{\text{true}}$ from the source datasets. For missing fields, the predicted values $\hat{y}_k$ from the Stage 1 model are used as replacements. Thus, the value for any field $k$ in the final composite sample $\hat{Y}$ is determined by the following equation.
\begin{equation}
\hat{y}_k =
\begin{cases}
y_k^{\text{true}}, & \text{if } y_k^{\text{true}} \text{ exists}, \\
\mathrm{Decode}\left(\mathcal{M}_{\text{stage1}}, P_k, x\right), & \text{otherwise}.
\end{cases}
\end{equation}

This iterative completion mechanism enables efficient synthesis of 3.6 million Pseudo-data samples, significantly expanding the joint training corpus.

\subsubsection{Data Quality Filtering and Consistency Verification}
Since the completed data contain fields generated by the Stage 1 model, they may include noise or logical inconsistencies. To improve data reliability, we apply a quality control protocol. First, we perform format validation and discard samples that cannot be parsed into valid JSON objects. Second, we conduct cross-task consistency verification by applying logical constraints between subtasks. For example, entities extracted in the QT field should be consistent with the corresponding expressions in the translation field.

After data cleaning, we merge the filtered pseudo data $\mathcal{D}_{\text{pseudo}}^{\text{filtered}}$, containing 3.6 million entries, with 310,000 high-quality human-labeled samples $\mathcal{D}_{\text{human}}$. The resulting mixture forms the final Stage 2 training dataset $\mathcal{D}_{\text{OneQuery}}$:
\begin{equation}
\mathcal{D}_{\text{OneQuery}} = \mathcal{D}_{\text{pseudo}}^{\text{filtered}} \cup \mathcal{D}_{\text{human}}
\end{equation}
where $|\mathcal{D}_{\text{pseudo}}^{\text{filtered}}| = 3.6 \times 10^6$ and $|\mathcal{D}_{\text{human}}| = 3.1 \times 10^5$. This hybrid dataset combines the broad coverage of large-scale pseudo data with the reliability of human-labeled data for OneQuery SFT.

In Stage 2, we use a unified OneQuery prompt format for supervised fine-tuning. To reduce distribution shift from Stage 1, this prompt keeps the instruction wording and field definitions as consistent as possible with the single-task prompts. We initialize the model with the Stage 1 weights and continue fine-tuning it to directly generate the complete JSON output for each input query. Training on the hybrid dataset enables the model to learn the unified output format and model semantic correlations among subtasks.

\subsection{Stage 3: RL Alignment}
Although Stage 2 SFT enables the model to generate unified OneQuery structures, it relies on token-level likelihood supervision over the complete JSON output. This may cause the model to learn surface-level shortcuts from the training data, such as over-relying on high-frequency lexical cues instead of the global semantics of the query. To mitigate this issue, we introduce RL in Stage 3 with task-aware reward signals. By penalizing outputs caused by such shortcuts, RL encourages the model to make predictions based on the overall query context and better align with the evaluation criteria of each subtask.

\subsubsection{RL Framework and Optimization Algorithm}
We employ GRPO, which updates the policy using relative advantages computed within a group of sampled outputs, thereby reducing the computational overhead of maintaining a separate value network.

For an input query \(x\), the policy model samples \(G\) distinct OneQuery outputs \(\{o_i\}_{i=1}^G\). Each output receives a scalar reward \(r_i\) from the multi-task rule-based reward function. Instead of training an additional value model as in PPO, GRPO normalizes rewards within the group to estimate the relative advantage:
\begin{equation}
A_i = \frac{r_i - \mu_r}{\sigma_r + \delta}
\end{equation}
where \(\mu_r\) and \(\sigma_r\) denote the mean and standard deviation of the group rewards, and \(\delta\) is a small constant for numerical stability. The optimization objective is:
\begin{equation}
\begin{aligned}
\mathcal{J}_{\text{GRPO}}(\theta) = \mathbb{E}\Big[ \frac{1}{G} \sum_{i=1}^{G} \min \big(& \rho_i(\theta) A_i, \\
& \mathrm{clip}(\rho_i(\theta), 1-\epsilon, 1+\epsilon) A_i \big) \Big]
\end{aligned}
\end{equation}
where the expectation is taken over \(x \sim \mathcal{D}\) and \(\{o_i\}_{i=1}^{G} \sim \pi_{\theta_{\text{old}}}\), and \(\rho_i(\theta) = \pi_\theta(o_i|x)/\pi_{\theta_{\text{old}}}(o_i|x)\) is the policy probability ratio. By optimizing this objective with task-aware rewards, the model is aligned across language recognition, spelling correction, query translation, and QT.

\subsubsection{Multi-Reward Function Design}
\label{sec:reward}

The reward function is critical for effective RL alignment. To match the multi-task nature of OneQuery, we design a composite reward \(R_{\text{total}}\) as a weighted sum of four subtask rewards:
\begin{equation}
R_{\text{total}} = w_1 R_{\text{lang}} + w_2 R_{\text{spell}} + w_3 R_{\text{trans}} + w_4 R_{\text{qt}}
\end{equation}
All rewards are rule-based and require no additional reward model training, providing transparent and reproducible feedback signals. We apply a normalization function \(\mathcal{N}(\cdot)\), including NFKC normalization, lowercasing, and whitespace normalization, before string comparison. The subtask rewards are defined as follows.

\paragraph{Language Recognition Reward}
Language recognition is evaluated by exact match between the predicted language tag and the reference label:
\begin{equation}
R_{\text{lang}} = \mathbb{I}\left[\mathcal{N}(\hat{y}_{\text{lang}}) = \mathcal{N}(y_{\text{lang}})\right]
\end{equation}

\paragraph{Spelling Correction Reward}
Spelling correction differs from the other subtasks because acceptable corrections may not be unique. A strict exact-match reward may assign zero credit to outputs that are close to the reference correction. We therefore use normalized edit similarity to provide partial credit when correction is needed.

Let \(x\) be the original query, \(y_{\text{spell}}\) be the reference correction, and \(\hat{y}_{\text{spell}}\) be the model prediction. We first determine whether correction is needed:
\begin{equation}
\text{NeedCorr} = \mathbb{I}\left[\mathcal{N}(x) \ne \mathcal{N}(y_{\text{spell}})\right].
\end{equation}
For brevity, let \(s=\mathcal{N}(\hat{y}_{\text{spell}})\), \(s^{*}=\mathcal{N}(y_{\text{spell}})\), and \(s_{0}=\mathcal{N}(x)\) denote the normalized prediction, reference correction, and original query, respectively. The spelling reward is defined as:
\begin{equation}
R_{\text{spell}} =
\begin{cases}
1.0, & \text{if } \text{NeedCorr}=1 \land s = s^{*} \\
\lambda \cdot \text{Edit}(s, s^{*}), & \text{if } \text{NeedCorr}=1 \land s \ne s^{*} \\
1.0, & \text{if } \text{NeedCorr}=0 \land s = s_{0} \\
0.0, & \text{otherwise}
\end{cases}
\end{equation}
where \(\lambda=0.3\) controls the maximum partial reward, and
\begin{equation}
\text{Edit}(a,b)=1-\frac{\text{LevDist}(\mathcal{N}(a),\mathcal{N}(b))}{\max(|\mathcal{N}(a)|,|\mathcal{N}(b)|)}
\end{equation}
is the normalized similarity based on Levenshtein edit distance. This design provides less sparse feedback while still assigning the highest reward only to exact corrections.

\paragraph{Translation Reward}
For query translation, we use exact match between the normalized predicted translation \(\hat{y}_{\text{trans}}\) and the reference translation \(y_{\text{trans}}\):
\begin{equation}
R_{\text{trans}} = \mathbb{I}\left[\mathcal{N}(\hat{y}_{\text{trans}}) = \mathcal{N}(y_{\text{trans}})\right]
\end{equation}
We use a binary reward because query translations are short and field-oriented, making soft metrics such as BLEU less reliable and potentially rewarding semantically shifted outputs.

\paragraph{Query Tagging Reward}
The output of the QT task is a structured set of labels. The prediction \(\hat{y}_{\text{qt}}\) and the reference \(y_{\text{qt}}\) are parsed into sets of tuples \(\hat{S} = \{(e_i, v_i)\}_{i=1}^{M}\) and \(S = \{(e_j, v_j)\}_{j=1}^{N}\), where \(e\) denotes the entity type and \(v\) denotes the entity text. The reward is computed as follows.
\begin{equation}
R_{\text{qt}} = \mathbb{I}\left[\hat{S} = S\right]
\end{equation}
This means the model receives a reward of 1.0 only if the predicted tuple set exactly matches the reference set, and 0.0 otherwise. For the same reason as in translation, we avoid partial-matching rewards such as F1.

\section{Experiments}

\subsection{Experimental Setup}

\paragraph{Evaluation Set}
We construct a unified OneQuery evaluation set \(\mathcal{D}_{\text{test}}\) consisting of 19,087 samples drawn from AliExpress online search logs, covering multilingual, multi-category real-world e-commerce queries. Each sample is manually annotated and verified by domain experts to provide ground-truth labels for all four fields. 

%精简了指标描述
\paragraph{Evaluation Metrics}
For language recognition, we use Accuracy, defined as the proportion of queries whose predicted language label exactly matches the reference. For query translation, we report ROUGE-1, ROUGE-2, and ROUGE-L to measure the lexical overlap between the model output and the reference translation, where ROUGE-L is computed based on the longest common subsequence. For QT, we use set-based Precision, Recall, and F1, where both predictions and annotations are parsed into sets of entity tuples and compared by exact tuple matching. Note that QT takes an English query as its input. In the pipeline and in Stage 1, the QT input is the output of the query translation step, so any translation error is directly passed to the tagging step. In contrast, from Stage 2 onwards OneQuery generates the QT field together with all the other fields from the original query within a single pass. For spelling correction, we compute Precision, Recall, and F1 by comparing the normalized predicted correction with the ground-truth. Before comparison, all text is processed with NFKC normalization.

\paragraph{Experimental Parameters}
All experiments are based on the e-commerce domain-continually pretrained model ME-LLM5.2-14B-Base. In the SFT stage, we utilize the AdamW optimizer with a learning rate of \(1 \times 10^{-4}\), a batch size of 128, and train for 1 epoch. In the RL stage, we employ the GRPO algorithm with a group sampling size \(G=8\), a KL penalty coefficient \(\beta=0.01\) (unless otherwise specified), a learning rate of \(5 \times 10^{-7}\), training for 1 epoch. The RL training data is randomly sampled from a pool of 310,000 Human-labeled OneQuery data, selecting 31,000 samples. All experiments are conducted on 16 $\times$ NVIDIA A100 80GB GPUs. Unless otherwise noted in Section~\ref{sec:reward_ablation}, all subsequent RL experiments in this paper use the optimal reward configuration by default.

%第一段修改了This section....
%倒数第二段加了不同规模模型比较，说明了模型泛化性
%倒数第一段加了闭源模型的比较，说明了闭源模型效果不好且自蒸馏是有必要的

\subsection{Main Results}
This section presents the comparison of our three-stage training pipeline with the independent single-task models, as shown in Table~\ref{tab:main_results}. The baseline is the multi-stage pipeline described in Section~\ref{sec:pipeline}. Besides the 14B backbone, we further report results on 8B and 4B backbones of the same domain-continually pretrained model, all following exactly the same three-stage procedure and hyperparameters. We also include two strong proprietary LLMs, Qwen3.8-Max~\cite{32} and GPT-5.6 Luna~\cite{34}, prompted with the same unified OneQuery instruction in a few-shot manner without any fine-tuning. Compared with the independent single-task baseline, our final Stage 3 model improves Language Recognition Acc from \(0.9514\) to \(0.9665\), Translation ROUGE-2 from \(78.20\) to \(80.50\), Spelling Correction F1 from \(0.8568\) to \(0.8717\), and QT F1 from \(0.8128\) to \(0.8366\).

\begin{table*}[t]
  \centering
  \small
  \caption{Main Experimental Results}
  \label{tab:main_results}
  \begin{tabular}{l l c ccc ccc ccc}
    \toprule
    \multirow{2}{*}{Model} & \multirow{2}{*}{Method} & Language & \multicolumn{3}{c}{Translation} & \multicolumn{3}{c}{Spelling Correction} & \multicolumn{3}{c}{QT} \\
    \cmidrule(lr){4-6} \cmidrule(lr){7-9} \cmidrule(lr){10-12}
    & & Acc & ROUGE-1 & ROUGE-2 & ROUGE-L & P & R & F1 & P & R & F1 \\
    \midrule
    \multirow{2}{*}{Few-Shot}
      & Qwen3.8-Max  & 0.9469 & 90.14 & 71.54 & 84.72 & 0.8050 & 0.8662 & 0.8345 & 0.6464 & 0.6344 & 0.6403 \\
      & GPT-5.6 Luna & 0.9505 & 88.06 & 66.16 & 80.92 & 0.7812 & 0.8913 & 0.8327 & 0.6207 & 0.6331 & 0.6268 \\
    \midrule
    \multirow{4}{*}{4B}
      & Pipeline & 0.9498 & 90.84 & 75.45 & 86.85 & 0.8658 & 0.8090 & 0.8365 & 0.8051 & 0.7979 & 0.8015 \\
      & Stage 1  & 0.9487 & 90.98 & 75.67 & 87.04 & 0.8593 & 0.8060 & 0.8318 & 0.8028 & 0.7959 & 0.7993 \\
      & Stage 2  & 0.9536 & 91.56 & 77.28 & 87.94 & 0.8863 & 0.8120 & 0.8475 & 0.8159 & 0.8084 & 0.8122 \\
      & Stage 3  & 0.9655 & 91.71 & 78.15 & 88.46 & 0.8907 & 0.7998 & 0.8428 & 0.8261 & 0.8213 & 0.8237 \\
    \midrule
    \multirow{4}{*}{8B}
      & Pipeline & 0.9513 & 91.78 & 77.26 & 87.91 & 0.8787 & 0.8285 & 0.8529 & 0.8111 & 0.8041 & 0.8076 \\
      & Stage 1  & 0.9493 & 91.73 & 77.39 & 87.98 & 0.8756 & 0.8195 & 0.8466 & 0.8112 & 0.8038 & 0.8075 \\
      & Stage 2  & 0.9538 & 91.56 & 77.30 & 87.94 & 0.8857 & 0.8120 & 0.8473 & 0.8159 & 0.8088 & 0.8123 \\
      & Stage 3  & 0.9662 & 92.69 & 80.45 & 89.77 & 0.9072 & 0.8019 & 0.8513 & 0.8394 & 0.8302 & 0.8348 \\
    \midrule
    \multirow{4}{*}{14B}
      & Pipeline & 0.9514 & 92.10 & 78.20 & 88.46 & 0.8851 & 0.8303 & 0.8568 & 0.8164 & 0.8093 & 0.8128 \\
      & Stage 1  & 0.9505 & 92.17 & 78.24 & 88.48 & 0.8856 & 0.8318 & 0.8578 & 0.8180 & 0.8109 & 0.8144 \\
      & Stage 2  & 0.9547 & 92.15 & 78.53 & 88.68 & 0.9024 & \textbf{0.8360} & 0.8679 & 0.8186 & 0.8132 & 0.8159 \\
      & Stage 3  & \textbf{0.9665} & \textbf{92.83} & \textbf{80.50} & \textbf{89.80} & \textbf{0.9134} & 0.8336 & \textbf{0.8717} & \textbf{0.8406} & \textbf{0.8326} & \textbf{0.8366} \\
    \bottomrule
  \end{tabular}
\end{table*}

We observe that the independent single-task baseline performs comparably to Stage 1, which suggests that jointly training the four subtasks within a single shared model does not degrade single-task performance.

From Stage 1 to Stage 2, OneQuery achieves notable improvements in Spelling Correction and QT, where Spelling Correction F1 rises from \(0.8578\) to \(0.8679\) and QT F1 from \(0.8144\) to \(0.8159\). These gains can be attributed to the large-scale synthetic pseudo data introduced in Stage 2, which increases data diversity and covers a broader range of spelling error patterns and query tagging cases. This helps alleviate the coverage limitations of the original single-task training data.

From Stage 2 to Stage 3, RL further improves performance across all tasks. This trend is consistent with our task-aware reward design, where different reward functions are tailored to the characteristics of different subtasks. Language Recognition obtains the largest improvement, from \(0.9547\) to \(0.9665\), partly because its performance decreases in Stage 2. In Stage 3, assigning a higher reward weight to this subtask (\(w_1 = 0.35\)) provides stronger optimization pressure and helps recover its performance.

For Translation, query outputs are short and field-oriented, where soft metrics such as BLEU can be unstable and may reward semantically shifted outputs. We therefore use a strict exact-match reward as a more conservative alternative, under which ROUGE-2 improves from \(78.53\) to \(80.50\) in Stage 3. For QT, the F1 score increases progressively across the three stages, with a gain
of \(2.07\) points in Stage 3 over Stage 2. We analyze the source of this
improvement in Section~\ref{sec:synergy}.

Finally, for Spelling Correction, the edit-distance-based reward provides denser feedback when the model partially corrects an error, and F1 further increases from \(0.8679\) to \(0.8717\) in Stage 3. This reward design encourages the model to make more cautious decisions between correcting the query and preserving the original input.

\paragraph{Generalization across Model Scales}
To examine whether the proposed framework depends on a particular model capacity, we repeat the entire three-stage procedure on 8B and 4B backbones with exactly the same data and hyperparameters. The relative trend remains highly consistent across all three scales. For Language Recognition, Stage 3 improves over the pipeline baseline by \(1.51\), \(1.49\), and \(1.57\) points on 14B, 8B, and 4B respectively; for QT, the F1 gains are \(2.38\), \(2.72\), and \(2.22\) points; and for Translation, ROUGE-2 increases by \(2.30\), \(3.19\), and \(2.70\) points. Spelling Correction is the only exception, where the smaller backbones do not exceed their own pipeline baselines, indicating that this subtask is the most sensitive to model capacity. Overall, these results show that the three-stage pipeline generalizes well and is not tied to a single model size.

\paragraph{Comparison with Few-Shot LLMs}
We further compare OneQuery with two strong proprietary LLMs, Qwen3.8-Max and GPT-5.6 Luna, prompted in a few-shot manner with the same unified OneQuery instruction. Both models perform reasonably well on Language Recognition, but fall far behind on QT, reaching only \(0.6403\) and \(0.6268\) in F1, about \(20\) points below the 14B Stage 3 model. A similar gap appears in Translation, where ROUGE-2 reaches only \(71.54\) and \(66.16\) compared with \(80.50\). On Spelling Correction they obtain the highest Recall but clearly lower Precision, indicating a tendency to over-rewrite queries that are already correct. In fact, OneQuery outperforms both proprietary LLMs on all metrics except Spelling Correction Recall at every backbone size, showing that domain-specific post-training matters more than model scale for this task. This observation also justifies our data synthesis design in Stage 2. A natural alternative would be to prompt a stronger general-purpose LLM to annotate the missing fields, but the results above show that such annotations would be unreliable, especially for QT, where a general LLM is unaware of the platform tag inventory. Self-distillation from the Stage 1 model is therefore not merely a cost-saving choice but a necessary one.

\subsection{Effectiveness Verification of the Data Synthesis Strategy}
To validate the effectiveness of the proposed pseudo-data synthesis plus human-labeled data construction strategy, we compare the performance of the joint SFT model under different data configurations, as shown in Table~\ref{tab:data_synthesis}.

\begin{table*}[t]
  \centering
  \small                                    % ← 加这一行
  \caption{Performance Comparison under Different Data Construction Configurations}
  \label{tab:data_synthesis}
  \begin{tabular}{l c ccc ccc ccc}
    \toprule
    \multirow{2}{*}{Dataset Composition} & Language & \multicolumn{3}{c}{Translation} & \multicolumn{3}{c}{Spelling Correction} & \multicolumn{3}{c}{QT} \\
    \cmidrule(lr){3-5} \cmidrule(lr){6-8} \cmidrule(lr){9-11}
    & Acc & ROUGE-1 & ROUGE-2 & ROUGE-L & P & R & F1 & P & R & F1 \\
    \midrule
    Human-labeled data only & \textbf{0.9644} & 89.41 & 73.61 & 85.90 & \textbf{0.9250} & 0.8010 & 0.8585 & \textbf{0.8222} & \textbf{0.8144} & \textbf{0.8183} \\
    Pseudo-data only & 0.9551 & \textbf{92.30} & 78.52 & 88.65 & 0.8857 & 0.8351 & 0.8597 & 0.8139 & 0.8086 & 0.8112 \\
    Human-labeled + Pseudo-data & 0.9547 & 92.15 & \textbf{78.53} & \textbf{88.68} & 0.9024 & \textbf{0.8360} & \textbf{0.8679} & 0.8186 & 0.8132 & 0.8159 \\
    \bottomrule
  \end{tabular}
\end{table*}

Compared with the SFT baseline trained only on human-labeled data, adding synthesized pseudo data notably improves Translation and Spelling Correction. For Translation, ROUGE-2 increases from \(73.61\%\) to \(78.53\%\), and ROUGE-L increases from \(85.90\%\) to \(88.68\%\). This suggests that the large-scale multilingual pseudo data generated by Iterative Field Completion helps alleviate the coverage limitations of the human-labeled translation subset. For Spelling Correction, F1 rises from \(0.8585\) to \(0.8679\), indicating that the diverse noise patterns in the synthesized pseudo data help the model recognize a wider range of spelling errors. For QT, the mixed-data model reaches an F1 of \(0.8159\), slightly lower than the human-only result of \(0.8183\). This suggests that synthetic QT annotations improve coverage but may also introduce noise that limits precision gains. For Language Recognition, the human-only model performs slightly better, likely because language classification has relatively simple patterns and the human-labeled set already provides sufficient coverage.

\subsection{RL Ablation Analysis}

\subsubsection{Reward Function Design Ablation}
\label{sec:reward_ablation}

The reward function plays an important role in RL alignment. To analyze the effect of different reward designs, we compare four configurations:
\begin{itemize}
    \item \textbf{reward-v1}. Uses binary exact-match rewards for Language Recognition and Spelling Correction, while using mean ROUGE for Translation and F1 for QT.
    \item \textbf{reward-v2}. Applies uniform binary Exact Match (0/1) reward to all subtasks.
    \item \textbf{reward-v3}. On the basis of reward-v2, replaces the binary Spelling Correction reward with a continuous score based on edit distance similarity \(r_{\text{edit}} = 1 - d_{\text{edit}}(\hat{y}, y) / \max(|\hat{y}|, |y|)\).
    \item \textbf{reward-v4}. Builds on reward-v3 by replacing the uniform weights with dependency-aware non-uniform weights \((w_1, w_2, w_3, w_4) = (0.35, 0.25, 0.25, 0.15)\). Since all fields are generated autoregressively, a field generated earlier influences a larger number of subsequent fields. We therefore give the largest weight to Language Recognition and the smallest to QT.
\end{itemize}

These RL experiments start from the Stage 2 model baseline, comparing the effects of the four defined reward functions. The results are presented in Table~\ref{tab:reward_ablation}.

\begin{table*}[t]
  \centering
  \caption{RL Performance Comparison under Different Reward Configurations}
  \label{tab:reward_ablation}
  \begin{tabular}{l c ccc ccc ccc}
    \toprule
    \multirow{2}{*}{Config} & Language & \multicolumn{3}{c}{Translation} & \multicolumn{3}{c}{Spelling Correction} & \multicolumn{3}{c}{QT} \\
    \cmidrule(lr){3-5} \cmidrule(lr){6-8} \cmidrule(lr){9-11}
    & Acc & ROUGE-1 & ROUGE-2 & ROUGE-L & P & R & F1 & P & R & F1 \\
    \midrule
    reward-v1 & 0.9643 & 91.72 & 78.96 & 88.12 & 0.8985 & 0.8294 & 0.8626 & 0.8296 & 0.8253 & 0.8274 \\
    reward-v2 & 0.9649 & 91.58 & 78.81 & 87.95 & 0.9012 & 0.8315 & 0.8650 & 0.8286 & 0.8205 & 0.8246 \\
    reward-v3 & 0.9644 & 92.05 & 79.43 & 88.51 & 0.8938 & 0.8221 & 0.8566 & 0.8314 & 0.8243 & 0.8278 \\
    reward-v4 & \textbf{0.9665} & \textbf{92.83} & \textbf{80.50} & \textbf{89.80} & \textbf{0.9134} & \textbf{0.8336} & \textbf{0.8717} & \textbf{0.8406} & \textbf{0.8326} & \textbf{0.8366} \\
    \bottomrule
  \end{tabular}
\end{table*}

Among the four reward designs, reward-v1 directly uses the evaluation metric of each subtask as its reward, but produces the weakest overall performance. One possible reason is that, under GRPO's intra-group normalization mechanism, continuous reward signals lack sufficient intra-group distinguishability. When the F1 scores of samples within the same group are clustered in a narrow range, inter-sample reward differences are dominated by sampling noise. After intra-group normalization, such noise is relatively amplified, reducing the signal-to-noise ratio of advantage estimation and making the policy-gradient direction unstable, thereby weakening the overall optimization effect.

reward-v2 replaces continuous rewards with sparse and high-contrast signals to improve intra-group sample distinguishability. Under this design, Spelling Correction F1 slightly increases from \(0.8626\) to \(0.8650\), while Query QT F1 decreases marginally from \(0.8274\) to \(0.8246\). These small and mixed changes suggest that simply replacing continuous rewards with sparse, high-contrast signals has inconsistent effects across tasks and brings limited overall gains. Meanwhile, the Translation results remain nearly unchanged.

Compared with reward-v2, reward-v3 improves both Translation and QT, with QT F1 increasing from \(0.8246\) to \(0.8278\), but Spelling Correction F1 drops to \(0.8566\). One possible explanation is that the reward signals across different tasks have different scales. During joint optimization, the contribution of each task to the total gradient becomes imbalanced, causing the Spelling Correction task to be relatively suppressed. This indicates that merely adjusting the form of reward signals is insufficient to ensure stable multi-task optimization; it is also necessary to further balance the gradient contributions across tasks.

reward-v4 is designed to address the above imbalance in gradient contributions by introducing non-uniform task weights. Specifically, a larger weight is assigned to Language Recognition (\(w_1=0.35\)) to compensate for the performance regression observed after Stage 2, while the weight of QT is reduced (\(w_4=0.15\)) to better balance its contribution in the weighted multi-task objective. After these adjustments, reward-v4 achieves the best performance on all four metrics among the evaluated designs. Overall, these results show that task-aware, fine-grained reward design, especially a reasonable balance of gradient contributions across tasks, is beneficial for multi-task reinforcement learning alignment.

\subsubsection{KL Penalty Coefficient Ablation Experiment}
The KL penalty coefficient \(\beta\) controls how strongly the RL-trained policy is constrained toward the SFT reference model. Under the reward-v4 configuration, we compare different values of \(\beta\), and the results are reported in Table~\ref{tab:kl_ablation}. When \(\beta=0.01\), the model achieves the best overall performance across the four metrics, suggesting that this setting provides a suitable balance between allowing policy improvement and preventing excessive deviation from the reference distribution.

\begin{table*}[t]
  \centering
  \caption{RL Performance Comparison under Different KL Penalty Coefficients}
  \label{tab:kl_ablation}
  \begin{tabular}{l c ccc ccc ccc}
    \toprule
    \multirow{2}{*}{Config} & Language & \multicolumn{3}{c}{Translation} & \multicolumn{3}{c}{Spelling Correction} & \multicolumn{3}{c}{QT} \\
    \cmidrule(lr){3-5} \cmidrule(lr){6-8} \cmidrule(lr){9-11}
    & Acc & ROUGE-1 & ROUGE-2 & ROUGE-L & P & R & F1 & P & R & F1 \\
    \midrule
    $\beta=0.1$ & 0.9652 & 92.41 & 79.87 & 89.15 & 0.9078 & 0.8295 & 0.8672 & 0.8290 & 0.8219 & 0.8254 \\
    $\beta=0.01$ & \textbf{0.9665} & \textbf{92.83} & \textbf{80.50} & \textbf{89.80} & \textbf{0.9134} & \textbf{0.8336} & \textbf{0.8717} & \textbf{0.8406} & \textbf{0.8326} & \textbf{0.8366} \\
    \bottomrule
  \end{tabular}
\end{table*}

In contrast, setting \(\beta=0.1\) introduces a stronger KL constraint and leads to consistent performance degradation. Specifically, compared with \(\beta=0.01\), Language Recognition accuracy decreases by \(0.13\) points, Translation ROUGE-2 by \(0.63\) points, Spelling Correction F1 by \(0.45\) points, and QT F1 by \(1.12\) points, all in absolute terms. These results indicate that an overly large KL penalty may overly restrict the policy update, making the model stay too close to the SFT reference model and thereby limiting the benefit of RL-based optimization. Therefore, we adopt \(\beta=0.01\) as the default setting.

\subsection{Case Study on Inter-Task Synergy and Error Tolerance}
\label{sec:synergy}
To further explain the two observations in the main experiments, namely how the unified paradigm handles error accumulation and why RL brings additional gains over joint SFT, we analyze two representative cases from the evaluation set. Table~\ref{tab:case_study} summarizes the fields generated by the two models on these cases. Incorrect results are underlined.

\begin{table*}[t]
  \centering
  \small
  \caption{Representative cases comparing the baseline with OneQuery}
  \label{tab:case_study}
  \begin{tabular}{l l c l l}
    \toprule
    Query & Model & Lang & Translation & Query Tagging \\
    \midrule
    \multirow{2}{*}{\textit{digitalizar video 8mm}}
      & Baseline & es & \underline{scan 8mm video} & FUNCTION \underline{scan}, FUNCTION video, MEASUREMENT 8mm \\
      & OneQuery & es & \underline{scan 8mm video} & FUNCTION digitize, FUNCTION video, MEASUREMENT 8mm \\
    \midrule
    \multirow{2}{*}{\textit{escape yamaha mt 07}}
      & Baseline & \underline{en} & \underline{escape yamaha mt 07} & \underline{MODEL escape}, MAIN\_BRAND yamaha, MODEL mt 07 \\
      & OneQuery & es & yamaha mt 07 exhaust & BRAND yamaha, MODEL mt 07, PRODUCT exhaust \\
    \bottomrule
  \end{tabular}
\end{table*}

The first case is the Spanish query \textit{digitalizar video 8mm}, where the user is looking for a service or device to digitize 8mm video tapes. Both the Independent Single-Task SFT baseline and our Stage 3 model translate it as \textit{scan 8mm video}, although \textit{digitize 8mm video} would be more appropriate. The baseline then predicts the key query term as \textit{scan}, showing that the translation error is directly propagated to the tagging result.

By contrast, our Stage 3 model still predicts the key query term as \textit{digitize}, even though its translation is also imperfect. This suggests that the unified model can use the original query and the overall context when generating query terms, rather than relying only on the generated translation. As a result, the translation error does not necessarily affect the tagging output, indicating that the unified formulation helps mitigate error accumulation.

This effect is reflected in the QT metrics of
Table~\ref{tab:main_results}. On the 14B backbone, QT F1 improves from \(0.8128\)
for the Traditional QU pipeline to \(0.8366\) for our Stage 3 model, with Precision
rising from \(0.8164\) to \(0.8406\) and Recall from \(0.8093\) to \(0.8326\).
Since QT is the last module of the pipeline and therefore absorbs the errors of all
upstream subtasks, the consistent gain on both Precision and Recall supports the
view that generating all fields jointly from the original query reduces the impact
of upstream errors.

The second case is the query \textit{escape yamaha mt 07}. The baseline identifies it as English, as \textit{escape} is a frequent English word and SFT model tends to associate its surface form with the English label. Since the pipeline invokes translation only for non-English queries, this incorrect language causes translation to be skipped and the original query to be passed to the tagging module. A single error in the language field thus affects multiple downstream fields.

Our Stage 3 model instead identifies the query as Spanish, as \textit{yamaha} and \textit{mt 07} indicate a motorcycle-accessory intent, where \textit{escape} refers to an exhaust pipe in Spanish rather than the English verb. A possible reason is that RL corrects the surface-form shortcut learned in SFT, so the model aggregates query-level signals instead of being misled by a single frequent token. As a result, \textit{escape} is correctly translated with the right product tag.

\subsection{Online Experiment Analysis}
To validate the industrial effectiveness of OneQuery, we deploy it on AliExpress using an offline-online scheme. The 14B OneQuery model performs offline inference over large-scale query logs. For head queries, the offline results are cached and served online directly; for the remaining tail queries, a lightweight 0.3B student model, distilled from the 14B OneQuery outputs, performs real-time inference. This design delivers strong query understanding while satisfying strict online latency constraints.

We conduct A/B tests across multiple countries, comparing OneQuery against the existing online query-understanding system. OneQuery improves the list-to-payment conversion rate (L2P) by 0.5\% and the relevance measured by side-by-side (SBS) human evaluation by 0.8 pt, and these gains remain stable throughout the experiment period. The results indicate that the more accurate and consistent query understanding produced by OneQuery benefits downstream retrieval and ranking, confirming the practical value of our framework under real-world traffic.

%增加了不同大小的sizes都有提升这句话
\section{Conclusion}
We present OneQuery, a unified query understanding framework for multilingual e-commerce search that jointly handles language recognition, spelling correction, translation, and query tagging within a single generation task. With supervised fine-tuning and reinforcement learning alignment, OneQuery achieves strong performance across multiple query understanding tasks. Meanwhile, OneQuery obtains consistent gains over the multi-stage pipeline across 4B, 8B, and 14B backbones, and clearly outperforms strong proprietary LLMs prompted in a few-shot manner. Experiments and online deployment on AliExpress further demonstrate its effectiveness and practical value for large-scale industrial search.

\section{Ethical Considerations}
All queries used in this work are collected from e-commerce search logs in accordance with the platform's terms of service, and contain no user identifiers or other personally identifiable information.

\begin{acks}
% Acknowledgements section remains empty per instruction
\end{acks}


\begin{thebibliography}{52}


\bibitem{2} Perez-Martin, J., Gomez-Robles, J., Guti\'errez-Fandi\~no, A., Adsul, P., Rajanala, S., and Lezcano, L. (2023). Cross-Lingual Search for E-Commerce Based on Query Translatability and Mixed-Domain Fine-Tuning. In Companion Proceedings of the ACM Web Conference 2023, pages 892--898.

\bibitem{4} Li, Y., Ma, S., Wang, X., Huang, S., Jiang, C., Zheng, H., Xie, P., Huang, F., and Jiang, Y. (2024). EcomGPT: Instruction-Tuning Large Language Models with Chain-of-Task Tasks for E-Commerce. In Proceedings of the AAAI Conference on Artificial Intelligence, volume 38, pages 18582--18590.

\bibitem{5} Peng, B., Ling, X., Chen, Z., Sun, H., and Ning, X. (2024). eCeLLM: Generalizing Large Language Models for E-Commerce from Large-Scale, High-Quality Instruction Data. arXiv preprint arXiv:2402.08831.

\bibitem{6} Ahmadvand, A., Kallumadi, S., Javed, F., and Agichtein, E. (2020). JointMap: Joint Query Intent Understanding for Modeling Intent Hierarchies in E-Commerce Search. In Proceedings of the 43rd International ACM SIGIR Conference on Research and Development in Information Retrieval, pages 1509--1512.

\bibitem{9} Xu, A., Lin, B., Xue, B., Wang, B., Xu, B., Wu, B., Zhang, B., Lin, C., Dong, C., Ling, C., et al. (2026). DeepSeek-V4: Towards Highly Efficient Million-Token Context Intelligence. arXiv preprint arXiv:2606.19348.

\bibitem{10} Yang, A., Li, A., Yang, B., Zhang, B., Hui, B., Zheng, B., Yu, B., Gao, C., Huang, C., Lv, C., et al. (2025). Qwen3 Technical Report. arXiv preprint arXiv:2505.09388.

\bibitem{12} Singh, A., Fry, A., Perelman, A., Tart, A., Ganesh, A., El-Kishky, A., McLaughlin, A., Low, A., Ostrow, A., Ananthram, A., et al. (2025). OpenAI GPT-5 System Card. arXiv preprint arXiv:2601.03267.

\bibitem{13} Metzler, D., Tay, Y., Bahri, D., and Najork, M. (2021). Rethinking Search: Making Domain Experts out of Dilettantes. ACM SIGIR Forum, 55(1):1--27.

\bibitem{15} Devlin, J., Chang, M., Lee, K., and Toutanova, K. (2019). BERT: Pre-Training of Deep Bidirectional Transformers for Language Understanding. In Proceedings of the 2019 Conference of the North American Chapter of the Association for Computational Linguistics: Human Language Technologies, Volume 1, pages 4171--4186.

\bibitem{16} Liu, Y., Ott, M., Goyal, N., Du, J., Joshi, M., Chen, D., Levy, O., Lewis, M., Zettlemoyer, L., and Stoyanov, V. (2019). RoBERTa: A Robustly Optimized BERT Pretraining Approach. arXiv preprint arXiv:1907.11692.

\bibitem{17} Sculley, D., Holt, G., Golovin, D., Davydov, E., Phillips, T., Ebner, D., Chaudhary, V., Young, M., Crespo, J., and Dennison, D. (2015). Hidden Technical Debt in Machine Learning Systems. In Advances in Neural Information Processing Systems, volume 28.

\bibitem{19} Behdin, K., Song, Q., Vasudevan, S., Sheng, J., Ma, X., Zhou, Z., Zhu, C., Li, G., Nguyen, C., Ghosh, S., et al. (2026). Scaling Up Large Language Models Serving Systems for Semantic Job Search. Proceedings of Machine Learning and Systems, 8:469--481.

\bibitem{21} Carmel, D. and Yom-Tov, E. (2010). Estimating the Query Difficulty for Information Retrieval. Morgan \& Claypool Publishers.

\bibitem{22} Raffel, C., Shazeer, N., Roberts, A., Lee, K., Narang, S., Matena, M., Zhou, Y., Li, W., and Liu, P. J. (2020). Exploring the Limits of Transfer Learning with a Unified Text-to-Text Transformer. Journal of Machine Learning Research, 21(140):1--67.

\bibitem{23} Chen, B., Guo, X., Wang, S., Liang, Z., Lv, Y., Ma, Y., Xiao, X., Xue, B., Zhang, X., Yang, Y., et al. (2025). OneSearch: A Preliminary Exploration of the Unified End-to-End Generative Framework for E-Commerce Search. arXiv preprint arXiv:2509.03236.

\bibitem{24} Dong, C., Yao, S., Jiao, P., Yang, J., Jin, Y., Huang, Z., Zhou, X., Ou, D., Tang, H., and Zheng, B. (2025). TaoSR1: The Thinking Model for E-Commerce Relevance Search. arXiv preprint arXiv:2508.12365.

\bibitem{25} Minaee, S., Mikolov, T., Nikzad, N., Chenaghlu, M., Socher, R., Amatriain, X., and Gao, J. (2024). Large Language Models: A Survey. arXiv preprint arXiv:2402.06196.

\bibitem{26} Huang, J., Huang, X., Zhao, F., Liu, Y., Zhang, H., Shi, F., Li, C., Sun, Z., Wu, Y., Hu, Y., et al. (2026). QP-OneModel: A Unified Generative LLM for Multi-Task Query Understanding in Xiaohongshu Search. arXiv preprint arXiv:2602.09901.

\bibitem{27} Liu, P., Shen, Q., Shen, J., Yao, C., Kao, K., Arora, R., Xu, D., Zheng, B., Ren, Y., Le, B., et al. (2026). A Unified Structured Query Understanding Framework for Industrial Semantic Search. arXiv preprint arXiv:2605.27441.

\bibitem{28} Wang, J., Dong, Z., Bai, B., Jiang, G., Yuan, A., and Cao, G. (2025). FoodGPT: Reinforcement Post-Training of Large Language Models in the Food Delivery Domain. In Proceedings of the 31st ACM SIGKDD Conference on Knowledge Discovery and Data Mining, Volume 2, pages 4966--4974.


\bibitem{30} Shao, Z., Wang, P., Zhu, Q., Xu, R., Song, J., Bi, X., Zhang, H., Zhang, M., Li, Y. K., Wu, Y., et al. (2024). DeepSeekMath: Pushing the Limits of Mathematical Reasoning in Open Language Models. arXiv preprint arXiv:2402.03300.

\bibitem{31} Schulman, J., Wolski, F., Dhariwal, P., Radford, A., and Klimov, O. (2017). Proximal Policy Optimization Algorithms. arXiv preprint arXiv:1707.06347.

\bibitem{32} Qwen Team. (2026). Qwen3.8-Max: A New Bar for Coding and Cowork. Technical Report, Alibaba Group.

\bibitem{34} OpenAI. (2026). GPT-5.6 System Card. Technical Report, OpenAI.


\bibitem{36} Guo, D., Yang, D., Zhang, H., Song, J., Wang, P., Zhu, Q., Xu, R., Zhang, R., Ma, S., Bi, X., et al. (2025). DeepSeek-R1: Incentivizing Reasoning Capability in LLMs via Reinforcement Learning. arXiv preprint arXiv:2501.12948.

\bibitem{37} Conneau, A., Khandelwal, K., Goyal, N., Chaudhary, V., Wenzek, G., Guzm\'an, F., Grave, E., Ott, M., Zettlemoyer, L., and Stoyanov, V. (2020). Unsupervised Cross-Lingual Representation Learning at Scale. In Proceedings of the 58th Annual Meeting of the Association for Computational Linguistics, pages 8440--8451.

\bibitem{38} Liu, P., Shen, J., Shen, Q., Yao, C., Kao, K., Xu, D., Arora, R., Zheng, B., Johnson, C., Hong, L., et al. (2025). Powering Job Search at Scale: LLM-Enhanced Query Understanding in Job Matching Systems. In Proceedings of the 34th ACM International Conference on Information and Knowledge Management, pages 4971--4975.

\bibitem{39} Peng, W., Li, G., Jiang, Y., Wang, Z., Ou, D., Zeng, X., Xu, D., Xu, T., and Chen, E. (2024). Large Language Model Based Long-Tail Query Rewriting in Taobao Search. In Companion Proceedings of the ACM Web Conference 2024, pages 20--28.

\bibitem{40} Cao, J., Li, S., Yu, B., Guo, X., Liu, T., and Wang, B. (2023). Towards Universal Cross-Domain Recommendation. In Proceedings of the Sixteenth ACM International Conference on Web Search and Data Mining, pages 78--86.

\bibitem{41} Desai, S., Rokon, M. O. F., Acharya, J. N., Shah, I., Yao, H., Porwal, U., and Lee, K. (2026). Unified Supervision for Walmart's Sponsored Search Retrieval via Joint Semantic Relevance and Behavioral Engagement Modeling. arXiv preprint arXiv:2604.07930.

\bibitem{42} Yang, Z., Pang, T., Feng, H., Wang, H., Chen, W., Zhu, M., and Liu, Q. (2024). Self-Distillation Bridges Distribution Gap in Language Model Fine-Tuning. In Proceedings of the 62nd Annual Meeting of the Association for Computational Linguistics (Volume 1: Long Papers), pages 1028--1043.

\bibitem{45} Ouyang, L., Wu, J., Jiang, X., Almeida, D., Wainwright, C., Mishkin, P., Zhang, C., Agarwal, S., Slama, K., Ray, A., et al. (2022). Training Language Models to Follow Instructions with Human Feedback. In Advances in Neural Information Processing Systems, volume 35, pages 27730--27744.

\bibitem{46} Wei, J., Wang, X., Schuurmans, D., Bosma, M., Ichter, B., Xia, F., Chi, E., Le, Q., and Zhou, D. (2022). Chain-of-Thought Prompting Elicits Reasoning in Large Language Models. In Advances in Neural Information Processing Systems, volume 35, pages 24824--24837.

\bibitem{48} Haque, M. A., Mehdi, A., Mahboob, M., and Fatima, T. (2026). Mitigating LLM Hallucinations through Domain-Grounded Tiered Retrieval. arXiv preprint arXiv:2603.17872.

\bibitem{49} Wang, L., Yang, N., and Wei, F. (2023). Query2doc: Query Expansion with Large Language Models. In Proceedings of the 2023 Conference on Empirical Methods in Natural Language Processing, pages 9414--9423.

\end{thebibliography}
\end{document}